\newcommand{\mystyleflag}{0}

\ifnum\mystyleflag=0
\documentclass[12pt]{article} 
\pdfoutput=1
\usepackage{jcappub}
\else
\documentclass[journal,article,submit,pdftex,moreauthors]{Definitions/mdpi} 
\firstpage{1} 
\pubvolume{1}
\issuenum{1}
\articlenumber{0}
\pubyear{2025}
\copyrightyear{2025}
\datereceived{ } 
\daterevised{ } 
\dateaccepted{ } 
\datepublished{ } 
\fi

\usepackage{amsmath}
\usepackage{amssymb}
\usepackage{physics}
\usepackage{slashed}
\usepackage{subfigure}
\usepackage{float}
\usepackage{notoccite}
\usepackage{cancel}
\def\be{\begin{equation}}
\def\ee{\end{equation}}
\def\bea{\begin{eqnarray}}
\def\eea{\end{eqnarray}}

\ifnum\mystyleflag=0
\title{Geometric framework for biological evolution}
\author[1,2]{Vitaly Vanchurin} 
\emailAdd{vitaly.vanchurin@gmail.com}
\affiliation[1]{Artificial Neural Computing, Weston, Florida, 33332, USA}
\affiliation[2]{Duluth Institute for Advanced Study, Duluth, Minnesota, 55804, USA}
\begin{document}
\else
\title{Geometric framework for biological evolution}
\Author{Vitaly  Vanchurin $^{1,2}$}
\AuthorNames{Vitaly  Vanchurin}
\AuthorCitation{Vanchurin, V.}
\address{%
$^{1}$ \quad Artificial Neural Computing, Weston, Florida, 33332, USA\\
$^{2}$ \quad Duluth Institute for Advanced Study, Duluth, Minnesota, 55804, USA}
\fi

\abstract{We develop a generally covariant description of evolutionary dynamics that operates consistently in both genotype and phenotype spaces. We show that the maximum entropy principle yields a fundamental identification between the inverse metric tensor and the covariance matrix, revealing the Lande equation as a covariant gradient ascent equation. This demonstrates that evolution can be modeled as a learning process on the fitness landscape, with the specific learning algorithm determined by the functional relation between the metric tensor and the noise covariance arising from microscopic dynamics. While the metric (or the inverse genotypic covariance matrix) has been extensively characterized empirically, the noise covariance and its associated observable (the covariance of evolutionary changes) have never been directly measured. This poses the experimental challenge of determining the functional form relating metric to noise covariance.
}

\ifnum\mystyleflag=0
\maketitle  
\else
\keyword{Geometric learning, Efficient learning, Emergent quantumness, Biological evolution} 
\begin{document}   
\fi

\section{Introduction}\label{sec:intro}

Theoretical modeling of evolutionary dynamics has a rich history spanning more than a century, from the foundational work in population genetics \cite{fisher1930genetical, wright1931evolution, haldane1927selection} to modern quantitative frameworks that describe how populations change under selection \cite{hartl2007principles, crow1970introduction, lynch2007origins, koonin2011logic}. Traditional approaches include population genetics models that track allele frequencies statistically \cite{fisher1930genetical, wright1931evolution, haldane1927selection} and quantitative genetics models that describe phenotypic response to selection through dynamical equations \cite{lande1976natural, lande1979quantitative,arnold1983morphology}. Despite their successes, these frameworks do not provide a coordinate-independent description of evolutionary dynamics, limiting their ability to capture the geometric and statistical structures underlying evolutionary processes. Even approaches that explicitly invoke geometric metaphors, such as fitness landscape theory \cite{wright1932roles,gavrilets2004fitness}, lack the differential geometric structure needed for a generally covariant formulation.

A recent paradigm shift has emerged from recognizing deep connections between evolutionary dynamics and learning theory \cite{vanchurin2022toward}. This framework builds on earlier work proposing that the universe on its most fundamental level can be understood as a neural network \cite{vanchurin2020world, vanchurin2021theory}. Under the evolution-as-learning framework, natural selection and replication arise naturally from the existence of a loss function (the negative of fitness) that is minimized during learning, and in sufficiently complex systems, the same learning phenomena occur on multiple scales. Subsequent work \cite{vanchurin2022thermodynamics} extended these ideas to the thermodynamics of evolution and the origin of life, revealing deep connections between learning theory, statistical mechanics, and evolutionary dynamics. The emergence of scale invariance in learning systems \cite{katsnelson2023emergent,kukleva2025dataset} further supports the claim that learning dynamics may be responsible for the multi-level structures observed across physical and biological systems.

More recently a covariant description of learning in machine learning systems was developed \cite{guskov2025covariant} and then generalized \cite{vanchurin2025geometric} as a unified geometric framework for describing learning in physical, biological, and machine learning systems. This work revealed three fundamental regimes characterized by the power-law relationship \(g \propto \kappa^{\alpha}\) between the metric tensor \(g\) and the noise covariance matrix \(\kappa\): the quantum regime (\(\alpha = 1\)), the efficient learning regime (\(\alpha = 1/2\)), and the equilibration regime (\(\alpha = 0\)). The emergence of the intermediate regime \(\alpha = 1/2\) was conjectured to be a key mechanism underlying biological complexity, and possibly the origin of life, suggesting that evolution may have discovered optimization algorithms more sophisticated than simple gradient descent. The present paper develops these ideas further by showing that the metric tensor on genotype space is naturally identified with the inverse genotypic covariance matrix through the maximum entropy principle \cite{jaynes1957information,jaynes1957information2}, transforming the Lande equation into a covariant gradient ascent equation of evolutionary dynamics (or covariant gradient descent equation of learning dynamics). This geometric perspective reveals that the specific learning algorithm implemented by biological evolution is determined by the functional relation \(g(\kappa)\) between the metric and noise covariance. While the metric has been extensively characterized empirically, the noise covariance remains unmeasured, posing an open challenge for evolutionary biology.

The paper is organized as follows. Section \ref{sec:genphen} introduces the genotype and phenotype spaces and the maps between them. Section \ref{sec:geometry} develops the geometric structures, including the pullback metric from phenotype to genotype space. Section \ref{sec:statistics} applies the maximum entropy principle to establish the identification between the inverse metric tensor and genotypic covariance. Section \ref{sec:lande} derives the Lande equation from the Price equation. Section \ref{sec:learning} connects evolution to learning algorithms and introduces possible functional relations \(g(\kappa)\). Section \ref{sec:experiment} discusses empirical observations of genotypic covariance spectra and the challenge of measuring noise covariance. The appendices provide detailed derivations of higher-order corrections and the relation between the fitness Hessian and noise covariance.

\section{Genotype and phenotype}\label{sec:genphen}

Consider a population of organisms, each characterized by its genetic sequence:
\begin{equation}
  \mathbf{s} = (s^1, s^2, \dots, s^K)
\end{equation}
where $s^\alpha \in \mathcal{A}$ represents the allelic state at locus $\alpha$, and $\mathcal{A}$ is a finite set of possible alleles with $|\mathcal{A}| = D$. For nucleotide sequences, $\mathcal{A} = \{A, C, G, T\}$ and $D = 4$. To enable the use of geometric and statistical structures, we define an embedding map:
\begin{equation}
    \hat{\mathbf{q}}: \mathcal{A}\rightarrow \mathbb{R}^d \label{eq:geno}
\end{equation}
where $d < D$ is the embedding dimension. The dimension $d$ can be chosen based on biological considerations, such as the number of independent chemical properties of nucleotides or through dimensionality reduction techniques applied to genomic data. The embedding map \eqref{eq:geno} allows us to assign to each discrete genotype $\mathbf{s}$ a continuous coordinate:
\begin{equation}
\mathbf{q}(\mathbf{s}) = \left( q^{\alpha 1}(\mathbf{s}), q^{\alpha 2}(\mathbf{s}), \dots, q^{\alpha d}(\mathbf{s}) \right) = \left( \hat{q}^1(s^\alpha), \hat{q}^2(s^\alpha), \dots, \hat{q}^d(s^\alpha) \right),
\end{equation}
where for each locus $\alpha$, the embedding coordinates are given by the map $\hat{\mathbf{q}}$ applied to the allelic state $s^\alpha$. The embedding is applied independently to each locus, which is why we use the same notation for all loci and coordinates.

Each discrete genotype gives rise to observable phenotypic traits through a phenotype map:
\begin{equation}
\hat{\mathbf{x}}: \mathcal{A}^K \rightarrow \mathbb{R}^N \label{eq:pheno}
\end{equation}
defined on the discrete space:
\begin{equation}
\hat{\mathbf{x}}(\mathbf{s}) = (\hat{x}^1(\mathbf{s}), \hat{x}^2(\mathbf{s}), \dots, \hat{x}^N(\mathbf{s})),
\end{equation}
where $N$ is the number of phenotypic traits considered. This map represents the biological processes through which genetic information produces observable characteristics; it also depends on the environment, which will be modeled later in the paper.

To leverage the continuous structure introduced by the embedding \eqref{eq:geno} and phenotype \eqref{eq:pheno} maps, we assume that $\hat{\mathbf{x}}$ can be interpolated to define a smooth map from a continuous genotype space to a continuous phenotype space:
\begin{equation}
  \mathbf{x}: \mathbb{R}^{dK}\rightarrow \mathbb{R}^N,
\end{equation}
such that for all discrete genotypes $\mathbf{s} \in \mathcal{A}^K$,
\begin{equation}
  \mathbf{x}(\hat{\mathbf{q}}(\mathbf{s})) = \hat{\mathbf{x}}(\mathbf{s}).
\end{equation}
This interpolation is not unique and must be chosen based on biological considerations, but its existence allows us to extend the phenotype map to continuous genotype values and compute derivatives such as $\partial x^i / \partial q^{\alpha r}$, which capture how continuous phenotype coordinates respond to infinitesimal changes in the continuous genotype coordinates.

Throughout this paper, we use the Einstein summation convention over repeated indices, with Latin indices $i,j,\dots$ running over phenotypic traits $1,\dots,N$, Greek indices $\alpha,\beta,\dots$ running over loci $1,\dots,K$, and Latin indices $r,s,\dots$ running over embedding coordinates $1,\dots,d$.

\section{Geometric structures}\label{sec:geometry}

The phenotype space $\mathbb{R}^N$ is equipped with a metric $G_{ij}(\mathbf{x})$ that quantifies distances between phenotypes. The squared distance between two infinitesimally close phenotypes $\mathbf{x}$ and $\mathbf{x} + d\mathbf{x}$ is:
\begin{equation}
ds^2 = G_{ij}(\mathbf{x}) \, dx^i dx^j.
\end{equation}
The inverse of this metric tensor is denoted with upper indices $G^{ij}(\mathbf{x})$, satisfying $G^{ik}G_{kj} = \delta^i_j$. A key feature of the geometric formulation is that the metric tensor $G_{ij}(\mathbf{x})$ is generally position-dependent, meaning that the geometry of phenotype space can vary with the phenotype itself. This allows the framework to capture phenomena where the relationships between traits change across different regions of phenotype space.

The interpolated phenotype map $\mathbf{x}: \mathbb{R}^{dK} \rightarrow \mathbb{R}^N$ allows us to pull back the metric from phenotype space to genotype space. For a genotypic displacement $dq^{\alpha r}$, the corresponding phenotypic displacement is 
\be
dx^i = \frac{\partial x^i}{\partial q^{\alpha r}} \, dq^{\alpha r}.
\ee 
The squared distance in genotype space, as measured by the induced metric, becomes:
\begin{align}
ds^2 &= G_{ij}(\mathbf{x}(\mathbf{q})) \, dx^i dx^j = G_{ij}(\mathbf{x}(\mathbf{q})) \left( \frac{\partial x^i}{\partial q^{\alpha r}} dq^{\alpha r} \right) \left( \frac{\partial x^j}{\partial q^{\beta s}} dq^{\beta s} \right) \nonumber\\
&= \left( \frac{\partial x^i}{\partial q^{\alpha r}} G_{ij}(\mathbf{x}(\mathbf{q})) \frac{\partial x^j}{\partial q^{\beta s}} \right) dq^{\alpha r} dq^{\beta s} = g_{\alpha r,\beta s}(\mathbf{q}) \, dq^{\alpha r} dq^{\beta s},
\end{align}
where the pullback metric tensor on genotype space is:
\begin{equation}
g_{\alpha r,\beta s}(\mathbf{q}) = \frac{\partial x^i}{\partial q^{\alpha r}} G_{ij}(\mathbf{x}(\mathbf{q})) \frac{\partial x^j}{\partial q^{\beta s}}.\label{eq:pull}
\end{equation}
This metric $g$ is the fundamental geometric structure on genotype space: it defines distances between genotypes not in terms of their genetic composition per se, but in terms of how different their phenotypes are, as measured by the phenotype metric $G_{ij}$. Two genotypes are considered close if they produce similar phenotypes, regardless of their molecular differences. This captures the biological intuition that what matters for evolution is phenotypic variation, not genetic variation for its own sake. 

Importantly, the pullback metric $g_{\alpha r,\beta s}(\mathbf{q})$ inherits the position-dependence from the phenotype metric $G_{ij}(\mathbf{x}(\mathbf{q}))$, making it a function of the genotype $\mathbf{q}$. This means that the geometry of genotype space is generally curved and varies across the space, reflecting how the same genetic change can have different phenotypic consequences depending on the genetic background. This geometric perspective complements classical quantitative genetics approaches to studying the $G$-matrix \cite{lande1979quantitative,arnold2008understanding}, which describes the patterns of genetic covariance among traits. The inverse of this pullback metric is denoted with upper indices $g^{\alpha r,\beta s}(\mathbf{q})$, satisfying $g^{\alpha r,\gamma t} g_{\gamma t,\beta s} = \delta^{\alpha r}_{\beta s}$.

\section{Statistical structures}\label{sec:statistics}

The population can be described by an ensemble of genotypes, represented by a probability distribution over the finite set of discrete genotypes. Let $\rho(\mathbf{s})$ denote the probability of observing genotype $\mathbf{s} \in \mathcal{A}^K$, satisfying $\sum_{\mathbf{s} \in \mathcal{A}^K} \rho(\mathbf{s}) = 1$. Each genotype $\mathbf{s}$ corresponds to a point $\hat{\mathbf{q}}(\mathbf{s})$ in the continuous genotype space via the embedding map $\hat{\mathbf{q}}: \mathcal{A}^K \rightarrow \mathbb{R}^{dK}$, with coordinates $q^{\alpha r}(\mathbf{s})$. Each genotype also maps to a phenotype ${\bf x}(\hat{\mathbf{q}}(\mathbf{s}))$ in phenotype space.

To apply the maximum entropy principle \cite{jaynes1957information,jaynes1957information2}, we specify a local reference frame in genotype space by choosing a locally flat reference metric. At any point of interest, such as the mean genotype $\bar{\bf q}$, we can construct Riemannian normal coordinates $\tilde{q}^{\alpha r}$ in which the reference metric becomes Euclidean and the volume element locally simplifies to the ordinary Lebesgue measure:
\begin{equation}
dV = d^{dK}\tilde{q}.
\end{equation}
The entropy of a genotypic distribution that is concentrated near the mean:
\begin{equation}
S[\tilde{\rho}] = -\int \tilde{\rho}(\tilde{\mathbf{q}}) \ln \tilde{\rho}(\tilde{\mathbf{q}}) \, d^{dK}\tilde{q},
\end{equation}
is to be maximized subject to constraints on the mean genotype:
\begin{equation}
   \int \tilde{q}^{\alpha r} \, \tilde{\rho}(\tilde{\mathbf{q}}) \, d^{dK}\tilde{q} = \bar{q}^{\alpha r}
\end{equation}
and on the expected squared distance:
\begin{equation}
  \int \delta_{\alpha r,\beta s} (\tilde{q}^{\alpha r} - \bar{q}^{\alpha r})(\tilde{q}^{\beta s} - \bar{q}^{\beta s}) \, \tilde{\rho}(\tilde{\mathbf{q}}) \, d^{dK}\tilde{q} = \sigma^2.
\end{equation}
This isotropic constraint, combined with the maximum entropy principle, yields a distribution that is spherically symmetric in the local coordinates:
\begin{equation}
\tilde{\rho}(\tilde{\mathbf{q}}) = \frac{1}{Z} \exp\left( -\frac{\lambda}{2} \delta_{\alpha r,\beta s} (\tilde{q}^{\alpha r} - \bar{q}^{\alpha r})(\tilde{q}^{\beta s} - \bar{q}^{\beta s}) \right),
\end{equation}
where $\lambda$ is a Lagrange multiplier determined by $\sigma^2$.

The covariance matrix of the maximum entropy distribution in the local reference frame is proportional to the identity:
\begin{equation}
c^{\alpha r,\beta s} = \int (\tilde{q}^{\alpha r} - \bar{q}^{\alpha r})(\tilde{q}^{\beta s} - \bar{q}^{\beta s}) \, \tilde{\rho}(\tilde{\mathbf{q}}) \, d^{dK}\tilde{q} = \frac{\sigma^2}{dK} \, \delta^{\alpha r,\beta s}.
\end{equation}
By appropriately rescaling the coordinates, the covariance matrix can be set to the identity. In any other coordinate system, it is given by the inverse metric tensor, leading to the fundamental identification:
\begin{equation}
g^{\alpha r,\beta s} = c^{\alpha r,\beta s}. \label{eq:gc}
\end{equation}

The mean phenotype $\bar{\mathbf{x}}$ is then given by the phenotype map applied to the mean genotype:
\begin{equation}
\bar{\mathbf{x}} = \mathbf{x}(\bar{\mathbf{q}}).
\end{equation}
Assuming the phenotype map is sufficiently smooth, i.e. locally linear,
\begin{equation}
  {x}^i(\mathbf{q}) - \bar{x}^i = \frac{\partial {x}^i}{\partial {q}^{\alpha r}} ({q}^{\alpha r} - \bar{q}^{\alpha r}),
\end{equation} 
we obtain a relationship between the phenotypic covariance $C^{ij}$ and the genotypic covariance $c^{\alpha r,\beta s}$:
\begin{equation}
C^{ij} =\frac{\partial {x}^i}{\partial {q}^{\alpha r}} \frac{\partial {x}^j}{\partial {q}^{\beta s}} \, c^{\alpha r,\beta s},\label{eq:pushc}
\end{equation} 
where the derivatives are evaluated at the mean genotype $\bar{\mathbf{q}}$. At the same time, the pushforward of the inverse metric from genotype space to phenotype space is given by:
\begin{equation}
G^{ij} = \frac{\partial {x}^i}{\partial {q}^{\alpha r}} \frac{\partial {x}^j}{\partial {q}^{\beta s}} \, g^{\alpha r,\beta s}.\label{eq:pushg}
\end{equation}
Combining Eqs. \eqref{eq:gc}, \eqref{eq:pushc} and \eqref{eq:pushg}, we obtain a relation between the inverse metric and the covariance matrix in phenotype space:
\begin{equation}
G^{ij} = C^{ij}. \label{eq:GC}
\end{equation}

\section{Lande equation}\label{sec:lande}

Consider a population with distribution \(\rho({\bf x})\) in phenotype space that undergoes evolution from one generation with state \({\bf x}\) to the next generation with state \({\bf x}'({\bf x})\), described by Wrightian fitness \({\cal W}(\bf{x})\). The expected change in the average state can be expressed as
\begin{equation}
\frac{\langle {\cal W}({\bf x})  {\bf x}'({\bf x})\rangle}{\left\langle {\cal W}({\bf x}) \right\rangle}  - \langle {\bf x} \rangle  = \frac{\left\langle {\cal W}({\bf x}) {\bf x} \right\rangle - \left\langle {\cal W}({\bf x}) \right\rangle \left\langle {\bf x} \right\rangle}{\left\langle {\cal W}({\bf x}) \right\rangle} + \frac{\left\langle {\cal W}({\bf x}) \left ( {\bf x}'({\bf x}) - {\bf x}\right )\right\rangle}{\left\langle {\cal W}({\bf x}) \right\rangle},\label{eq:price}
\end{equation}
where expectations are defined with respect to the invariant volume element:
\begin{equation}
  \langle \cdots \rangle = \int \cdots \, \rho({\bf x}) \,\sqrt{\det G}\, d^N x.
\end{equation}
Equation \eqref{eq:price} is a continuous form of the Price equation \cite{price1970selection, price1972extension}, usually written as 
\begin{equation}
\Delta \langle {\bf x} \rangle  = \frac{\mathrm{Cov}({\cal W}({\bf x}),{\bf x} )}{\left\langle {\cal W}({\bf x}) \right\rangle} + \frac{\mathrm{E}\left ({\cal W}({\bf x}) \Delta {\bf x}\right )}{\left\langle {\cal W}({\bf x}) \right\rangle}\label{eq:price2}
\end{equation}
where \(\Delta \langle {\bf x} \rangle \equiv  \frac{\langle {\cal W}({\bf x})  {\bf x}'({\bf x})\rangle}{\left\langle {\cal W}({\bf x}) \right\rangle}  - \langle {\bf x} \rangle\) is the change in the mean phenotype from one generation to the next, and \(\Delta {\bf x} \equiv {\bf x}'({\bf x}) - {\bf x}\) is the individual-level phenotypic change between parent and offspring. Assuming perfect transmission, meaning that offspring resemble their parents on average (\(\mathrm{E}\left ({\cal W}({\bf x}) \Delta {\bf x}\right ) = 0\)), the Price equation simplifies to the selection response:
\begin{equation}
\Delta \langle {\bf x} \rangle  = \frac{\mathrm{Cov}({\cal W}({\bf x}),{\bf x} )}{\left\langle {\cal W}({\bf x}) \right\rangle}.\label{eq:selection}
\end{equation}

The Wrightian fitness \({\cal W}({\bf x})\) can be expanded around \(\bar{\bf x}\):
\begin{equation}
{\cal W}({\bf x}) = {\cal W}(\bar{\bf x}) + \frac{\partial {\cal W}}{\partial \bar{x}^i} (x^i - \bar{x}^i) + \mathcal{O}(|x-\bar{x}|^2),\label{eq:w_expand}
\end{equation}
where derivatives are evaluated at the mean phenotype \(\bar{\bf x}\). The mean fitness is then:
\begin{equation}
\langle {\cal W} \rangle = {\cal W}(\bar{\bf x}) + \frac{\partial {\cal W}}{\partial \bar{x}^i} \langle x^i - \bar{x}^i \rangle + \mathcal{O}(|x-\bar{x}|^2) = {\cal W}(\bar{\bf x}) + \mathcal{O}(|x-\bar{x}|^2).\label{eq:mean_w}
\end{equation}
Substituting the linear approximations \eqref{eq:w_expand} and \eqref{eq:mean_w} into the selection response \eqref{eq:selection} we get:
\begin{align}
\Delta \langle x^i \rangle &= \frac{\mathrm{Cov}({\cal W}({\bf x}), x^i)}{\langle {\cal W} \rangle} \approx \frac{1}{{\cal W}(\bar{\bf x})} \mathrm{Cov}\left( {\cal W}(\bar{\bf x}) + \frac{\partial {\cal W}}{\partial \bar{x}^j} (x^j - \bar{x}^j), x^i \right) \nonumber\\
&= \frac{1}{{\cal W}(\bar{\bf x})} \frac{\partial {\cal W}}{\partial \bar{x}^j} \, \mathrm{Cov}(x^j, x^i) = \frac{\partial \log {\cal W}}{\partial \bar{x}^j} \, C^{ij},\label{eq:cov_step}
\end{align}
where the phenotypic covariance matrix is 
\begin{equation}
  C^{ij} = \mathrm{Cov}(x^i, x^j) = \langle (x^i - \bar{x}^i)(x^j - \bar{x}^j) \rangle.
\end{equation}
Equation \eqref{eq:cov_step} is the discrete form of the Lande equation \cite{lande1976natural, lande1979quantitative}, usually written in continuous time as 
\begin{equation}
  \frac{d \bar{x}^i}{dt} = C^{ij}(\bar{\bf x}) \frac{\partial {\cal F}(\bar{\bf x})}{\partial \bar{x}^j} \label{eq:lande}
\end{equation}
where \({\cal F}(\bar{\bf x}) = \log {\cal W}(\bar{\bf x})\) is the Malthusian fitness evaluated at the mean phenotype. (See also Ref. \cite{vanchurin2025geometric} for an alternative derivation of the Lande equation.) The resulting equation describes the change in mean phenotype as a product of the genetic covariance matrix and the selection gradient, providing a foundation for understanding evolutionary dynamics in quantitative genetics. Higher-order nonlinear corrections to the Lande equation are derived in Appendix \ref{sec:higher}.

\section{Learning dynamics}\label{sec:learning}

In the previous sections, we established two types of structures: the geometric structures $G$ and $g$, and the statistical structures $c$ and $C$. By applying the maximum entropy principle, the phenotypic covariance matrix $C$ was identified with the inverse metric tensor $G^{-1}$ \eqref{eq:GC}, and thus the Lande equation \eqref{eq:lande} can be viewed as learning dynamics described by covariant gradient ascent \cite{guskov2025covariant}:
\begin{equation}
  \frac{d \bar{x}^i}{dt} = G^{ij}(\bar{\bf x}) \frac{\partial {\cal F}(\bar{\bf x})}{\partial \bar{x}^j},
\end{equation}
with the loss function taken to be the negative of Malthusian fitness. The dynamics can be pulled back to genotype space:
\begin{equation}
\frac{d \bar{q}^{\alpha r}}{dt} = g^{\alpha r,\beta s}(\bar{\mathbf{q}}) \, \frac{\partial {\cal F}({\mathbf{x}}(\bar{\mathbf{q}}))}{\partial \bar{q}^{\beta s}} = g^{\alpha r,\beta s}(\bar{\mathbf{q}}) \, \frac{\partial {\cal F}({\mathbf{x}})}{\partial {x}^i} \, \frac{\partial {x}^i}{\partial \bar{q}^{\beta s}},\label{eq:cga}
\end{equation}
where the inverse genotype metric $g^{-1}$ is identified with the genotype covariance matrix $c$ \eqref{eq:gc}. 

Note that fitness depends on the genotype coordinates $\mathbf{q}$ only implicitly through the phenotype map $\mathbf{x}(\mathbf{q})$, i.e., ${\cal F}(\bar{\mathbf{x}}(\bar{\mathbf{q}})) = {\cal F}(\bar{\mathbf{x}})$. This establishes a direct analogy with machine learning systems: the genotype $\mathbf{q}$ corresponds to trainable variables (weights and biases), while the phenotype $\mathbf{x}$ corresponds to non-trainable variables (neuron states). The fitness function depends on the phenotypes, which are determined by the genotypes through the phenotype map. Evolution thus optimizes fitness function by adjusting genotypic variables, just as learning algorithms optimize a loss function by updating weights and biases.

The microscopic counterpart of \eqref{eq:cga} can be modeled as \cite{vanchurin2025geometric}:
\begin{equation}
\frac{d q^{\alpha r}}{dt}
=
g^{\alpha r,\beta s}(\mathbf q)
\frac{\partial {\cal H}(\mathbf x(\mathbf q),\hat{\mathbf x})}{\partial q^{\beta s}},
\label{eq:cgd}
\end{equation}
where $\hat{\mathbf x}$ represents environmental degrees of freedom. The microscopic fitness is
\begin{equation}
{\cal H}(\mathbf x(\mathbf q),\hat{\mathbf x})
\approx
{\cal F}(\mathbf x(\mathbf q)) + \phi(\mathbf q,t),\label{eq:FHphi}
\end{equation}
with the stochastic component satisfying
\begin{align}
\langle \phi(\mathbf q,t) \rangle_\tau &= 0, \\
\langle \phi(\mathbf q,t)\phi(\mathbf q',t') \rangle_\tau
&=
{\cal C}(\mathbf q,\mathbf q')\,\delta(t-t'),
\end{align}
and the noise covariance given by
\begin{equation}
\kappa_{\alpha r,\beta s}(\mathbf q)
=
\left\langle
\frac{\partial\phi}{\partial q^{\alpha r}}
\frac{\partial\phi}{\partial q^{\beta s}}
\right\rangle
=
\frac{\partial}{\partial q^{\alpha r}}
\frac{\partial}{\partial q'^{\beta s}}
{\cal C}(\mathbf q,\mathbf q')
\Big|_{\mathbf q'=\mathbf q}.\label{eq:noise}
\end{equation}

A potential relation between the metric tensor, which is related to the genotypic covariance (see Section \ref{sec:statistics}), and the noise covariance, which is related to the fitness Hessian (see Appendix \ref{sec:hessian}), can be motivated by learning theory, where many efficient learning algorithms arise by selecting the metric tensor as a specific function of the noise covariance. A particularly simple class of such functions is the power-law dependence \cite{guskov2025covariant}:
\begin{equation}
g(\kappa) = \kappa^{a},\label{eq:power}
\end{equation}
where $a = 0$ corresponds to stochastic gradient descent (or ascent) and $a = 1$ corresponds to natural gradient descent (or ascent). More generally, efficient learning algorithms, such as Adam \cite{kingma2014adam} and AdaBelief \cite{zhuang2020adabelief},  effectively implement metrics of the form
\begin{equation}
g(\kappa) = \kappa^{1/2} + \epsilon I,\label{eq:adam}
\end{equation}
where $\epsilon$ is some constant. These algorithms belong to the broader class of covariant gradient descent methods, where the metric tensor adapts based on statistical information extracted from the gradient noise.

\section{Genotype statistics}\label{sec:experiment}

To develop a phenomenological model of the geometry of genotype space, we recall that the inverse metric tensor $g^{-1}$ is given by the genotypic covariance of the population of organisms \eqref{eq:gc}:
\begin{equation}
g^{\alpha r,\beta s}
=\left\langle
q^{\alpha r} q^{\beta s} 
\right\rangle - \left\langle
q^{\alpha r}\right\rangle \left\langle q^{\beta s} 
\right\rangle .
\label{eq:qcov}
\end{equation}
Empirical studies of genotypic covariance matrices have revealed that the eigenvalue spectrum exhibits a rapid decay, with the first few eigenvalues capturing a large fraction of the total genetic variance \cite{qin2018power}. The decay of the eigenvalues is often well approximated by a power law:
\begin{equation}
\lambda_i \propto i^{-\alpha},
\end{equation}
with exponent $\alpha$ typically in the range $1.0$ to $2.0$. For such a power law, the effective rank of the covariance matrix can be expressed in terms of the Riemann zeta function:
\begin{equation}
r_{\text{eff}}(\alpha) =
\frac{(\sum_i \lambda_i)^2}{\sum_i \lambda_i^2}
\approx
\frac{(\sum_i i^{-\alpha})^2}{\sum_i i^{-2\alpha}}
=
\frac{\zeta(\alpha)^2}{\zeta(2\alpha)},
\end{equation}
although the power-law dependence is usually modified at both large and small eigenvalues. Empirical values of the effective rank are often on the order of $\sim 10^2$, which is much smaller than the total genomic sequence length $\dim(g) \sim 10^9$. In practice, this means that to obtain the actual metric tensor (or the inverse of the genotype covariance matrix) we may need to use a pseudo-inverse to avoid problems associated with inverting zero or near-zero eigenvalues.

While substantial empirical data exist for the genotypic covariance matrix $g^{-1}$, direct empirical estimates of the noise covariance matrix $\kappa$ remain unavailable. Using the microscopic equation \eqref{eq:cgd}, the noise covariance \eqref{eq:noise} with raised indices can be expressed as the covariance of temporal changes:
\begin{equation}
\kappa^{\alpha r, \beta s} =
g^{\alpha r,\gamma t} \, \kappa_{\gamma t,\delta u} \, g^{\delta u, \beta s}
=
\left\langle
\frac{d q^{\alpha r}}{dt}
\frac{d q^{\beta s}}{dt}
\right\rangle
-
\left\langle
\frac{d q^{\alpha r}}{dt}
\right\rangle
\left\langle
\frac{d q^{\beta s}}{dt}
\right\rangle .
\label{eq:qdotcov}
\end{equation}
Estimating this matrix would require time-series data tracking evolutionary trajectories across many generations, combined with the ability to separate deterministic selection from stochastic fluctuations, a formidable empirical challenge that has yet to be met.

Given the observational data for the genotype covariance $g^{-1}$ \eqref{eq:qcov} and the raised-index noise covariance $g^{-1} \kappa g^{-1}$ \eqref{eq:qdotcov}, the metric tensor $g$ and the noise covariance $\kappa$ should be related through some function $g(\kappa)$ that describes the learning algorithm. For example, if the functional dependence is a power law \eqref{eq:power}, then
\begin{equation}
  g^{-1} = \left( g^{-1} \kappa g^{-1} \right)^{\frac{a}{2a-1}}.
\end{equation} 
For stochastic gradient ($a=0$) and natural gradient ($a=1$), we obtain $g^{-1} = I$ and $g^{-1} = g^{-1} \kappa g^{-1}$, respectively. For efficient learning algorithms such as those described by \eqref{eq:adam}, we obtain
\begin{equation}
  g^{-1} = \epsilon^{-1} \left( I - \sqrt{g^{-1} \kappa g^{-1}} \right),
\end{equation}
which also shows the significance of the $\epsilon$ parameter. However, the precise functional form of $g(\kappa)$, and therefore the specific learning algorithm implemented in biological evolution, remains an open question. This disconnect between theory and available data underscores the need for new experimental approaches aimed at directly characterizing not only the genotype covariance matrix $g^{-1}$, but also the covariance matrix of evolutionary changes of genotypes $g^{-1} \kappa g^{-1}$. 

\section{Discussion}\label{sec:discuss}

In this paper, we developed a geometric framework for biological evolution that unifies concepts from differential geometry, statistical mechanics, and learning theory. Three main results emerged from this analysis.

Firstly, we constructed a generally covariant formulation of evolutionary dynamics that operates consistently in both genotype and phenotype spaces. This is accomplished by embedding the discrete genotype space \(\mathcal{A}^K\) into a continuous space \(\mathbb{R}^{dK}\) via an embedding map \(\hat{\mathbf{q}}:\mathcal{A}\rightarrow\mathbb{R}^d\). The phenotype map \(\hat{\mathbf{x}}:\mathcal{A}^K\rightarrow\mathbb{R}^N\) is then interpolated to a smooth function \(\mathbf{x}:\mathbb{R}^{dK}\rightarrow\mathbb{R}^N\), allowing geometric structures to be pushed forward and pulled back between the two spaces. This framework provides a coordinate-invariant language for describing evolutionary processes, where all equations maintain their form under arbitrary reparameterizations of either space.

Secondly, we applied the maximum entropy principle to derive the fundamental identifications between the inverse metric and the covariance matrix in both genotype space, \(g^{\alpha r,\beta s} = c^{\alpha r,\beta s}\), and phenotype space, \(G^{ij} = C^{ij}\). These identifications are fully consistent with the Lande equation, which emerges as the leading-order contribution from an expansion of the Price equation. Higher-order corrections to the Lande equation, derived in Appendix~\ref{sec:higher}, introduce contributions from third cumulants (skewness) and reveal the conditions under which the linear approximation breaks down.

Thirdly, the identification of metric with covariance shows that the Lande equation is precisely a covariant gradient ascent equation, implying that biological evolution can be understood as a learning process on the (negative of) fitness landscape. The specific learning algorithm implemented by evolution is determined by the functional relation \(g(\kappa)\) between the metric tensor \(g\) and the noise covariance \(\kappa\) that appears in the microscopic dynamics. While the inverse metric (or the genotypic covariance matrix) \(g^{-1}\) has been extensively characterized empirically, the noise covariance \(\kappa\) and its associated observable \(g^{-1}\kappa g^{-1}\) (the covariance of evolutionary changes of genotypes) have never been directly reconstructed from observations.

The functional form of \(g(\kappa)\) determines whether evolution implements simple algorithms such as stochastic gradient descent \(g = I\), or natural gradient descent \(g = \kappa\), or more sophisticated algorithms like those used in modern machine learning optimizers \cite{guskov2025covariant}. Without empirical access to \(\kappa\), the true learning algorithm of evolution cannot be determined. This gap between theory and data defines a clear research program for evolutionary biology. Future experimental work should aim to characterize not only the static covariance of genotypes but also the dynamical covariance of genotypic changes. Such measurements would allow us to infer the functional relation \(g(\kappa)\) and thereby identify the specific learning algorithm that nature has discovered.

{\it Acknowledgments.} The author is grateful to Mikhail Katsnelson and Eugene Koonin for many stimulating discussions and comments on the manuscript.

\bibliographystyle{unsrt}
\bibliography{library}
\appendix 

\section{Higher-order corrections}\label{sec:higher}

The Lande equation \eqref{eq:lande} is a first-order approximation derived by linearizing the fitness function around the mean phenotype. To understand its limitations and obtain more accurate dynamics, we extend the expansion to second order.

Expanding the fitness function to second order around the mean phenotype gives:
\begin{equation}
{\cal W}({\bf x}) = {\cal W}(\bar{\bf x}) + \frac{\partial {\cal W}}{\partial \bar{x}^i} (x^i - \bar{x}^i) + \frac{1}{2} \frac{\partial^2 {\cal W}}{\partial \bar{x}^i \partial \bar{x}^j} (x^i - \bar{x}^i)(x^j - \bar{x}^j) + \mathcal{O}(|x-\bar{x}|^3)\label{eq:w_second}
\end{equation}
and then the mean fitness is:
\begin{equation}
\langle {\cal W} \rangle = {\cal W}(\bar{\bf x}) + \frac{1}{2} \frac{\partial^2 {\cal W}}{\partial \bar{x}^i \partial \bar{x}^j} C^{ij} + \mathcal{O}(|x-\bar{x}|^3).\label{eq:mean_second}
\end{equation}
The covariance between fitness and phenotype becomes:
\begin{equation}
\mathrm{Cov}({\cal W}({\bf x}), x^i) = \frac{\partial {\cal W}}{\partial \bar{x}^j} C^{ji} + \frac{1}{2} \frac{\partial^2 {\cal W}}{\partial \bar{x}^j \partial \bar{x}^k} S^{jki} + \mathcal{O}(|x-\bar{x}|^4),\label{eq:cov_second}
\end{equation}
where the third cumulant (skewness) is:
\begin{equation}
  S^{jki} = \langle (x^j - \bar{x}^j)(x^k - \bar{x}^k)(x^i - \bar{x}^i)\rangle.
\end{equation}
Substituting into \eqref{eq:selection} gives:
\begin{equation}
\Delta \langle x^i \rangle = \frac{1}{{\cal W}(\bar{\bf x})} \frac{\partial {\cal W}}{\partial \bar{x}^j} C^{ji} + \frac{1}{2{\cal W}(\bar{\bf x})} \frac{\partial^2 {\cal W}}{\partial \bar{x}^j \partial \bar{x}^k} S^{jki} + \mathcal{O}(|x-\bar{x}|^4).\label{eq:lande_skew}
\end{equation}
Converting to Malthusian fitness \({\cal F} = \log {\cal W}\) using \(\frac{\partial {\cal W}}{\partial \bar{x}^i} = {\cal W} \frac{\partial {\cal F}}{\partial \bar{x}^i}\) and keeping only the leading correction:
\begin{equation}
\frac{d\bar{x}^i}{dt} = C^{ij} \frac{\partial {\cal F}}{\partial \bar{x}^j} + \frac{1}{2} S^{ijk} \frac{\partial^2 {\cal F}}{\partial \bar{x}^j \partial \bar{x}^k}  + \mathcal{O}(|x-\bar{x}|^4).\label{eq:lande_higher}
\end{equation}
For symmetric or Gaussian distributions, the third cumulant vanishes (\(S^{ijk}=0\)), and the Lande equation \eqref{eq:lande} is recovered. These corrections vanish for symmetric distributions (zero skewness), in which case the Lande equation remains accurate to higher order.

\section{Fitness Hessian}\label{sec:hessian}

The metric tensor, which is also the genotypic covariance \eqref{eq:gc}, 
\begin{equation}
g^{\alpha r,\beta s}
=
\left\langle
(q^{\alpha r}-\bar q^{\alpha r})
(q^{\beta s}-\bar q^{\beta s})
\right\rangle ,\label{eq:gencov}
\end{equation}
can be differentiated to obtain:
\begin{equation}
\frac{d g^{\alpha r,\beta s}}{dt}
=
2 g^{\beta s,\gamma t}\frac{\partial^2 {\cal F}}
{\partial \bar q^{\gamma t}\partial \bar q^{\delta u}}
g^{\alpha r,\delta u}
+
\left\langle
g^{\alpha r,\gamma t}
\frac{\partial \phi}{\partial q^{\gamma t}}
(q^{\beta s}-\bar q^{\beta s})
\right\rangle
+
\left\langle
(q^{\alpha r}-\bar q^{\alpha r})
g^{\beta s,\gamma t}
\frac{\partial \phi}{\partial q^{\gamma t}}
\right\rangle ,
\end{equation}
where we used \eqref{eq:cgd} and kept only terms to second order in the expansion of ${\cal F}$.

Using the Furutsu-Novikov formula for Gaussian white noise, the correlation between noise gradient and deviation gives
\begin{equation}
\left\langle
\frac{\partial \phi}{\partial q^{\gamma t}}
(q^{\beta s}-\bar q^{\beta s})
\right\rangle
=
\frac{1}{2} g^{\beta s,\delta u} \kappa_{\gamma t,\delta u}.
\end{equation}
Combining all contributions, we obtain the evolution equation for the inverse metric:
\begin{equation}
\frac{d g^{\alpha r,\beta s}}{dt}
=
g^{\alpha r,\gamma t} \left( 2\frac{\partial^2 {\cal F}}
{\partial \bar q^{\gamma t}\partial \bar q^{\delta u}} + \kappa_{\gamma t,\delta u}\right) g^{\delta u,\beta s}.
\end{equation}
or for the metric itself:
\begin{equation}
\frac{d g_{\alpha r,\beta s}}{dt} = -2\frac{\partial^2 {\cal F}}
{\partial \bar q^{\alpha r}\partial \bar q^{\beta s}} - \kappa_{\alpha r,\beta s}.
\end{equation}
In the stationary limit where the metric (or genotype covariance \eqref{eq:gencov}) is static, we obtain the balance condition
\begin{equation}
\frac{\partial^2 {\cal F}}
{\partial \bar q^{\gamma t}\partial \bar q^{\delta u}}
=
- \frac{1}{2} \kappa_{\gamma t,\delta u}.\label{eq:Fkappa}
\end{equation}

\end{document}